\documentclass{aa}
\usepackage{graphicx}
\begin{document}
   \title{Dissipative Collapse of a Spherical Cluster of Gas
           Clouds} 
           
\author{K.Indulekha,
	\inst{1}
	  G.V.Vijayagovindan\inst{1} 
	 \and
	 S.Ramadurai\inst{2}
	 }
	\offprints{S.Ramadurai}
			 
       \institute{School of Pure and Applied Physics,
          Mahatma Gandhi University,
            Kottayam  686 560, India\\
         \and
	     Tata Institute of Fundamental Research, Homi Bhabha
 	      Road, Colaba, Mumbai 400 005, India\\
 	      \email{durai@tifr.res.in}
		}

\abstract{
 With a view to understand the galaxy/star formation scenario, we
  investigate the dissipative collapse of a spherical cluster of
gas clouds with an isotropic velocity distribution.  The time
scale for collapse to one tenth radius 
is studied as a function of the collision time
in the system. The scalar virial equation is used to investigate
the evolution of the size of the cluster. This is supplemented
with an evolution equation for the random kinetic energy.
The above system is numerically solved and the results
analyzed. For large values of the collision time we find that the time scale
for collapse is proportional to the collision time as expected.  However for
large values of the dissipation, i.e. for small collision times, 
 the collapse time shows a nonlinear dependence
on the collision time.  
             }

\date{Received       / Accepted      }
 
\authorrunning{K.Indulekha et. al.}
\titlerunning{Dissipative Collapse of Gas Clouds}   
\maketitle
%
%

\section{Introduction}

The formation of stars and galaxies normally proceeds from the 
collapse of gas clouds. Hence it is imperative that one investigates the 
evolution of the size and eccentricity of spherical and 
spheroidal clusters of gas clouds.  Such an investigation 
 for point masses with an isotropic distribution of velocities 
 was made by Chandrasekhar and Elbert (\cite{Chandrasekhar and Elbert 1972})
  using the scalar 
 and tensor virial equations.     
This was followed by Som Sunder and Kochhar 
(\cite{Som Sunder and Kochhar 1985}), who 
further  examined (\cite{Som Sunder and Kochhar 1986}) 
the evolution for anisotropic velocity
distributions also.  Here we examine the dissipative evolution of
a spherical cluster of  gas clouds with an isotropic
velocity distribution, using the scalar virial equation,   in an
attempt to obtain the time scale, for collapse impeded by
virialized mass motions.  The gravitational binding energy
released during collapse feeds the random motions of the clouds.
This kinetic energy is dissipated in
cloud-cloud collisions.  The evolution of the cluster size is
followed using the scalar virial equation.  It is supplemented
with an evolution equation for the kinetic energy of random
motions.   From this investigation we get quantitative 
relations between the collapse time and the collision time 
under various conditions.   
%

\section{The model}

        We consider a spherically symmetric cluster of $N$ 
equal mass clouds
of radius $R_c$ and mass $M_c$ distributed uniformly.  
The clouds have a one
dimensional r.m.s. velocity  $v_{rms}$. The gas clouds are in pressure balance with an
intercloud medium.   The mass in the intercloud medium is taken to
be too small to affect the dynamics and the contribution from it
to the virial as well as the dissipation  will be ignored
throughout (Mestel and Paris \cite{Mestel and Paris 1984}). 
  The scalar virial equation for
the system is (Chandrasekhar and Elbert
\cite{Chandrasekhar and Elbert 1972})

\begin{eqnarray}
\frac{1}{2}
\frac{d^{2}I}{dt^{2}}
=2T+\Omega
\end{eqnarray}

where $I=4\pi\int\rho(r)r^{4}dr$ is like the moment of inertia of
the system about  the centre and $\rho(r)$ is the density at a
distance $r$ from the origin. $T$  is the
kinetic energy associated with mass motions, and  $\Omega$ is the
self gravitational potential energy of the cluster. The evolution
of the system must be consistent with equation 1 and we may use it
to investigate the evolution of  the gross size of the cluster.
For  the
cluster we get
\begin{eqnarray}
I=\frac{3}{5}MR^{2}(1+(\frac{R_{c}}{R})^{2})
\end{eqnarray}
and

\begin{eqnarray}
\Omega=-\frac{3}{5}\frac{GM^{2}}{R}(1-5(\frac{R_{c}}{R})^{2}).
\end{eqnarray}
Thus for $R_{c}$ much smaller than R using the expressions for I and
$\Omega$ corresponding to a homogeneous distribution of matter would
not be too erroneous.

         The velocity of the clouds may be separated into
two parts, a random part and a mean motion.  The
kinetic energy for the random motions say $T_{R}$ is $\frac{3}{2}Mv_{rms}^{2}$. 
If we consider only homologous changes in the
size of the system with the mean velocity at a point being
proportional to the distance of the point from the centre,  we get 
the kinetic energy for the mean motion as 
$T_{M} =\frac{3}{10}M(\frac{dR}{dt})^{2}$.  The cluster is assumed to be
 uniform at all times. Introducing
\begin{eqnarray}
I=\frac{3}{5}MR^{2},  
T_{M}=\frac{3}{10}M(\frac{dR}{dt})^{2},  
\Omega=-\frac{3}{5}\frac{GM^{2}}{R},
\end{eqnarray}
and
\begin{eqnarray}
T_{R}=\frac{3}{2}Mv_{rms}^{2}
\end{eqnarray}

the virial equation becomes

\begin{eqnarray}
\frac{d^{2}R}{dt^{2}}= \frac{5v_{rms}^{2}}{R}-\frac{GM}{R^{2}}.
\end{eqnarray}

The evolution equation for the random kinetic energy will have a
growth term due to the feeding of gravitational potential energy
into the random motions and a dissipative term due to cloud
collisions. In the case of nonuniform collapse the gravitational
energy released during contraction  goes initially into radial
flow, but eventually some of it is converted into the energy of
random motions. The possibility that all the released binding
energy is converted into random kinetic energy  has also been
advocated (Odgers and Stewart \cite{Odgers and Stewart 1958}).
 We consider that all
the released potential energy is converted to random
kinetic energy. This gives

\begin{eqnarray}
\frac{dT_{R}}{dt}\mid_{growth}=-\frac{d\Omega}{dt}  
\end{eqnarray}

For dissipation by collisions the dissipation rate will be proportional
to (the number of collisions suffered by a cloud in unit time)
x (the energy loss per  collision) x (the total number of
clouds).  The number of collisions per unit time will be
proportional to the (geometrical cross section of the cloud) x
(the velocity of the cloud) x (the number density of clouds).
The dissipation rate is thus proportional to
$Mv_{rms}^{3}NR_{c}^{2}/R^{3}= $$\nu$$Mv_{rms}^{3}/R$.  Here we have
expressed $R_{c}$ in terms of the cloud filling factor f as
$R_{c}=(\frac{f}{N})^{\frac{1}{3}}R$. Putting  all the numerical 
factors into a parameter $a$  we write 
$\nu=aN^{\frac{1}{3}}f^{\frac{2}{3}}$.   Here $\nu$ is proportional 
to the number of collisions in a free fall time
 for virial theorem random motions in the 
 system. As an
example, Elmegreen(\cite{Elmegreen 1985}) when considering that the initial speed of 
a cloud is halved after interacting with a mass equal to itself 
via collisions, gets 
$a=\frac{9\sqrt{6}}{16}$.  We thus get

\begin{eqnarray}
\frac{dT_{R}}{dt}\mid_{loss}={-\nu}
\frac{Mv_{rms}^{3}}{R}
\end{eqnarray}

        Putting all the terms together we get the evolution 
equation for the random
kinetic energy $T_{R}$ in the form
\begin{eqnarray}
\frac{dv_{rms}^{2}}{dt}=\frac{-2GM}{5R^{2}}\frac{dR}{dt}-
{\nu}\frac{2v_{rms}^{3}}{3R}
\end{eqnarray}

As the  cluster contracts it is the contraction produced by the
kinematics of the collapse itself that mainly affects the
radius of the clumps rather than any self gravitational
effects Lacey(\cite{Lacey 1984}).  For clouds which are 
themselves Jeans' unstable the collapse time for the clouds will be
smaller than the  free fall collapse time for the cluster
only by a factor $\surd{f}$.    The number of the clouds can
vary  as there can be cloud disruptions or
coalescences also during collisions.  We assume a constant
$\nu$ throughout the evolution.

          For $v_{rms}^2$$\sim$$\frac{GM}{R_{0}}$ corresponding to virial
 equilibrium the collision time is  $\sim$$1/{\nu}$ in
 units of the free fall time $t_{ff}$$\sim$$\surd(\frac{R_{0}^{3}}{GM})$,
 where $R_{0}$ is the initial radius. 
Thus since dissipation is due to collisions the
 dissipation time is also $\sim$$1/{\nu}$.  For the system to be collisional the
mean free path of the clouds should be less than twice the radius
of the cloud. This yields the condition
$N^{\frac{1}{3}}f^{\frac{2}{3}}>1/6$.  This condition is independent of the
size of the cluster since we have assumed a constant filling factor throughout
the collapse.  
%

\section{Results}
 We have considered evolution of the cluster under two different conditions: 
(a) the system is initially in virial equilibrium with
$T_{R}=\Omega/2$   (b) the system has zero initial kinetic
energy. As case (c) we have considered the evolution of the cluster
 starting from virial equilibrium and with no gravitational feeding .
Equations (6) and (9) were solved for  various values of  the
only free parameter ${\nu}$$\sim$$1/t_{c}$ where $t_{c}$ is the collision time for
virial theorem random motions in units of $ t_{ff}$. 
 The equations were normalized and integrated 
with a constant time step.  Desired
accuracy was achieved by using a time step which was appropriately
small.  It was checked that with zero dissipation the change in
the total energy was less than one percent  over 10 units of normalized
time.

%
%
  \begin{figure}
   \resizebox{\hsize}{!}{\includegraphics{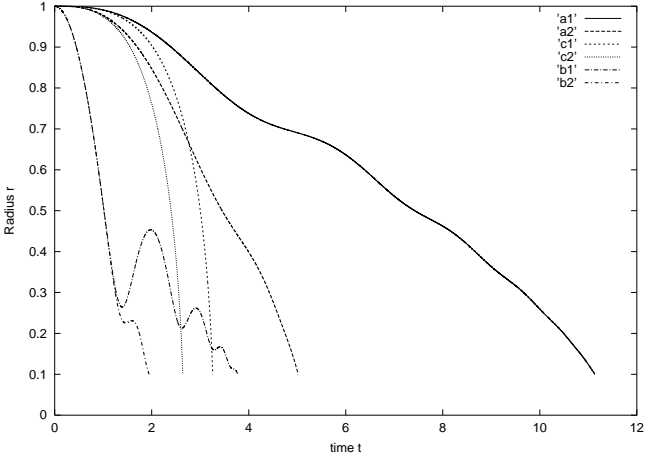}}
   \caption{R as a function of t for
    $t_{c}$=$5$ and $2$ for cases(a), (b) and (c) 
    with $t$ in units of $t_{ff}$. }
   \label{FigGam}%
   \end{figure}

In Fig. 1 we give R as a
function of time for the cases (a) and (b) for two different
 values of ${\nu}$, $.2$ and $0.5$.  The two curves with faint oscillations are
for case (a). The curves with strongly marked oscillations
 are for case (b) and the curves for no gravitational
 feeding, case (c) show monotonic
collapse.   For
larger  values of ${\nu}$ the oscillations start at smaller radii.

%
%

   \begin{figure}
   \resizebox{\hsize}{!}{\includegraphics{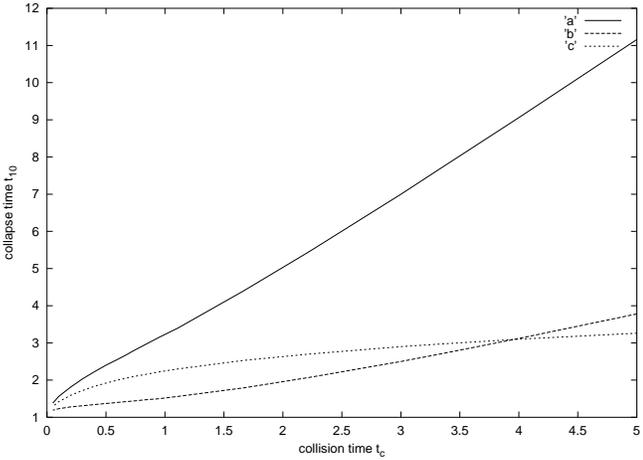}}
   \caption{$t_{10}$ as a function of $t_{c}$ for
  	 cases (a),(b) and (c).
  	  Both times are in units of $t_{ff}$.}
              \label{FigGam}%
    \end{figure}

In Fig. 2 we show $t_{10}$ the time for collapse
 to one-tenth
size in units of  $t_{ff}$ 
as a function of  $t_{c}(=(aN^{1/3}f^{2/3})^{-1})$,
 The uppermost 
curve is for case (a), the lowermost for
case (b), and the middle curve is for 
case (c). For large values of $t_{c}$, $t_{10}$
 depends linearly on $t_{c}$ and a fit to the curves 
in this region gives,  for case (a)

\begin{eqnarray}
t_{10}=0.95+2.0t_{c},
\end{eqnarray}

and for case (b) we get

\begin{eqnarray}
t_{10}=0.74+0.6t_{c}.
\end{eqnarray}

In the nonlinear region the curves can be represented 
with very good accuracy
by a quadratic in $t_{c}$.  For case (a) we get 

\begin{eqnarray}
t_{10}=1.25+3.0t_{c}-1.4t_{c}^{2}
\end{eqnarray}

and for case (b)

\begin{eqnarray}
t_{10}=1.16+0.63t_{c}-0.46t_{c}^{2}
\end{eqnarray}

Fig.2 also shows the case of no 
gravitational feeding (case (c)),
for comparison with earlier work where the dissipation of
virial theorem random motions was investigated by various authors.
  The cluster is
initially in virial equilibrium.   For small 
dissipation i.e. for large collision times $t_{10}$ is not very
 sensitive to $t_{c}$ and increases slowly   with
 $t_{c}$ in a linear fashion  as  $t_{10}=2.2+0.2t_{c}$. However 
for large dissipation, there is  a fast increase of the
 collapse time as the collision time increases. This nonlinear portion 
is fitted by the curve $t_{10}=1.2+2.1t_{c}-1.38t_{c}^{2}$.

%

\section{Discussion}

For a nondissipative system starting with zero kinetic energy
 Chandrasekhar and Elbert (\cite{Chandrasekhar and Elbert 1972}) 
 have shown that the system will execute
 oscillations.  If we put zero dissipation in our system we get the same
oscillations.  For the finite dissipation cases which we
 have considered we find these oscillations;
 however they are of a damped character due to the finite value of the
dissipation.  

Examining $R$ as a function of $t$, in all the cases 
with gravitational feeding we find that collapse takes
place in an oscillatory fashion.  The oscillations are well marked in  case (b).
  Here there is an initial  rapid collapse when $T_{R}$ rapidly builds up and
finally reaches a value high enough for the virial equilibrium condition to
hold.  This takes place when the value of $\frac{dR}{dt}$ is quite large and
the resulting bounce of the system is quite hard.  This causes the oscillations
to be strongly marked.  As $\nu$ increases the bounce is delayed due to the
 the slower increase of $T_{R}$.  In case (a) the bounce can take place quite
early on and is quite mild.   For large dissipation the bounce gets
 delayed so much that we do not get
it within collapse to one tenth size and the collapse looks apparently
monotonic.

   It is known that a self gravitating cloud
collapsing without dissipation will
satisfy the conditions for virial equilibrium when it has
collapsed to half radius (Hoyle \cite{Hoyle 1949}).
  The idea of
virialization of mass motions in gravitational collapse and their
subsequent dissipation has been considered by many people.(Odgers and
Stewart \cite{Odgers and Stewart 1958},
Larson \cite{Larson 1977},
Mestel \cite{Mestel 1977},
Scalo and Pumphrey \cite{Scalo and Pumphrey 1982},
Elmegreen \cite{Elmegreen 1985}.) 
These works consider virial theorem random
motions and the time scale for their dissipation. Simulations of
the collisional dissipation of virial theorem random motions by
Scalo and Pumphrey give a dissipation time $\sim$$20$ free fall times.
(See also Chieze and Lazareff \cite{Chieze and Lazareff 1980},
Hausman \cite{Hausman 1981}.)
 Elmegreen considered a magnetized cloud of clumps and obtained 
 dissipation times of the order $3-5$
cloud crossing times.

       We have analyzed collapse with gravitational feeding, 
of clusters starting from virial
equilibrium and also starting with zero kinetic energy (like for example from
turn around). For small values of the dissipation like 
say  $t_{c}=5t_{ff}$ we get $t_{10}$ about 
$11$ for case (a) and about $3.8$ for case (b).  In the  model of
Scalo and  Pumphrey (\cite{Scalo and  Pumphrey 1982}) 
the number of clouds get smaller with
time due to cloud coalescence.  The resultant slower dissipation
rate explains the comparatively longer dissipation times they get.
As a general statement we may say that for small dissipation, 
clusters starting from virial equilibrium can take more than twice the
dissipation time to collapse, when there is gravitational feeding. 
 With no
gravitational feeding we get $t_{10}>t_{ff}>t_{c}$.  The increase in the
collapse time as the collision time increases is less pronounced in this case
and for large collision times lies in the range $2-4$ free fall times.  This is
consistent with the results obtained by other authors for the dissipation time
for virial theorem random motions.  As $t_{c}$ goes to zero
$t_{10}$ rapidly approaches the value $1.1$ which is the time for free fall collapse to
one tenth size in the units we have chosen. Our results indicate that for large
collision times gravitational feeding can significantly increase the time
scales for the dissipative collapse of clusters of gas clouds.

For large values of $t_{c}$ i.e. small values of the dissipation we see from
fig. 1 that the collapse time depends linearly on $t_{c}$.  For case (a) this
may be understood as follows.  For small dissipation the cluster is collapsing
very slowly.  Thus growth as well as loss of $T_{R}$  is small 
and the cluster contraction
is governed mainly  by the decrease of $T_{R}$ in the virial equation.
 This decrease is
in proportion to $t_{c}$.  For case (b) also after an initial rapid collapse, when
$T_{R}$ rapidly builds up, the size of the cluster
is governed similarly by the decrease of $T_{R}$ in equation (6).  
For large dissipation  however the collapse being rapid, changes in 
$\frac{dR}{dt}$
 and  $R$ are large causing large changes in $T_{R}$.  There is thus a 
strong and nonlinear dependence of
the evolution of $T_{R}$ on $t_{c}$.  The collapse time also reflects 
 this strong 
 nonlinear dependence.

For the system under consideration the free fall time provides the
appropriate time scale for the system to adjust to changes in the
virial theorem (Brosche 1970).   While noticing that a quasi virial evolution 
does not hold for dissipation times shorter than the dynamical time, 
we find distinctive difference in the 
dependance of $t_{10}$ on $t_{c}$ in the two regions 
 $t_{c}>t_{ff}$  and 
 $t_{c}<t_{ff}$.

The consequences of this investigation for the formation of stars and galaxies
are being examined at present in a quantitative fashion and will be reported 
separately.

%

\begin{acknowledgements}

KI thanks Profs. Rajaram Nityananda, and Kandaswamy Subramanian of NCRA, Pune  
for helpful discussions. She also thanks IUCAA, Pune for an 
associateship during part of this investigation.  
\end{acknowledgements}

\end{document}